\documentclass[aps,prl,twocolumn,superscriptaddress,showpacs,showkeys]{revtex4}
\usepackage{epsfig}
\usepackage{graphicx}
\usepackage{amsmath}
\usepackage[english]{babel}

\begin{document}

\title{Explosive percolation in scale-free networks}

\author{Filippo Radicchi}
\affiliation{Complex
  Networks and Systems Group, ISI Foundation, Torino, Italy}
\author{Santo Fortunato}
\affiliation{Complex Networks and
  Systems Group, ISI Foundation, Torino, Italy}

\begin{abstract}

We study scale-free networks constructed via a cooperative Achlioptas growth process. Links
between nodes are introduced in order to produce a scale-free graph with given exponent $\lambda$
for the degree distribution, but the choice
of each new link depends on the mass of the clusters that this link will merge. Networks
constructed via this biased procedure show a percolation transition which strongly differs from
the one observed in standard percolation, where links are introduced just randomly.
The different growth process leads to a phase transition with a non-vanishing percolation
threshold already for $\lambda>\lambda_c\sim 2.2$.
More interestingly, the transition is
continuous when $\lambda \leq 3$ but becomes discontinuous when $\lambda > 3$. 
This may have important consequences both for the structure of networks and for the
dynamics of processes taking place on them.

\end{abstract}

\maketitle

The modern science of networks~\cite{Newman:2003,vitorep,barratbook} has opened new perspectives
in the study of complex systems. The simple graph representation, where the elementary units of a system
become nodes and their mutual interactions links connecting the nodes pairwise, enables one to understand a lot of properties about the 
structure and dynamics of a system. In particular, the degree distribution $P(k)$, i. e. the probability distribution of the 
number of neighbors $k$ of a node, plays an important role. Real networks often display skewed degree distributions,
where many nodes with low degree coexist with some nodes with high degree ({\it hubs}). The presence of the hubs
is responsible for a number of striking properties, like a high resilience against random failures/attacks~\cite{albert00}
and the absence of an epidemic threshold~\cite{pastor01}. Resilience is determined by checking what is the fraction
of nodes/links that need to be removed in order to split the network into a set of microscopic disjoint connected components.
This is closely related to the
process of {\it percolation}~\cite{staufferbook}, where one studies the conditions leading to the formation of 
a macroscopic ({\it giant}) component of the network. Here one starts from a set of nodes and no links; links are added randomly or 
according to a certain rule, until a giant component is formed. 
On networks having power law degree distributions (scale-free networks)
with exponent $\lambda$ smaller than $3$, 
the fraction of nodes/links to be removed from the graph for it to have no giant component tends to $1$
in the limit of infinite network size~\cite{cohen00}. In the spirit of percolation, and focusing on links, 
this can be read the other way around: a scale-free network with $\lambda<3$
is kept connected by a vanishing fraction of randomly chosen links, i.e. the percolation threshold is zero. For 
$\lambda>3$, instead, a finite threshold appears. Indeed, a giant component exists if the average number $z_2$ of next-to-nearest neighbors
of a node exceeds the average number $z_1=\langle k\rangle$ of its nearest neighbors~\cite{newman01}. On networks without
degree-degree correlations~\cite{pastor00}, $z_2=\langle k^2\rangle-\langle k\rangle$, which diverges when the exponent $\lambda$ of 
$P(k)$ is smaller than $3$, whereas it is finite when $\lambda>3$. 
The divergence of the variance $\langle k^2\rangle$ is generally a sufficient condition 
to ensure the absence of a percolation threshold on a scale-free graph, with or without degree-degree correlations~\cite{dorogovtsev08},
although in the case of large disassortativity a finite percolation threshold may emerge~\cite{vazquez04}.

In any case, whether there is a finite threshold or not, the percolation transition in networks is continuous: the order parameter,
represented by the relative size of the giant component with respect to the whole system, 
varies continuously from zero starting from the critical point. 
This is due to the fact that links are (usually) randomly placed on the network. 
Recent work by Achlioptas and coworkers has shown that, for networks similar to 
Erd\"os-R\'enyi random graphs~\cite{erdos}, the percolation transition becomes discontinuous (first-order)
if links are placed according to special non-random rules~\cite{achlioptas09}. Such growth processes for graphs are meanwhile known 
as {\it Achlioptas processes}, and the resulting connectedness transition as {\it explosive percolation}. 
Discontinuous transitions triggered by similar mechanisms 
were previously observed in the jamming of information packets on communication networks~\cite{echenique05}.  
In this letter we want to explore what happens if one grows a scale-free network via an Achlioptas growth 
process. We will see that the resulting scenario is very different than in the case of ordinary percolation.

\begin{figure}
\begin{center}
\includegraphics[angle=-90,width=\columnwidth]{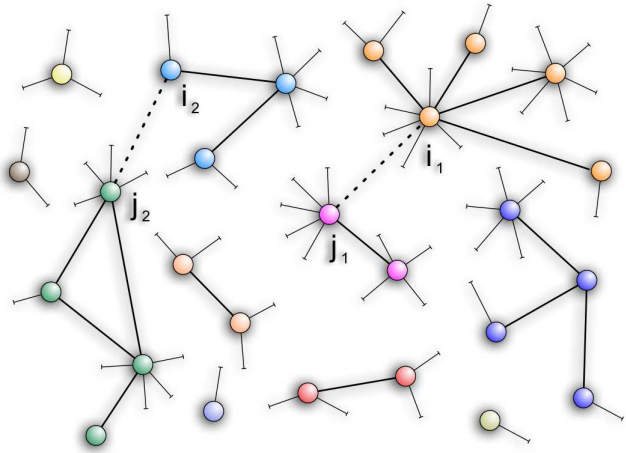}
\end{center}
\caption{Scheme of the construction process of a network via an Achlioptas process with product rule (PR). Two pairs of 
stubs are taken at random (each pair is indicated by the dotted lines), and the products of the sizes of 
each pair of clusters merged by joining the stubs are computed. The 
stubs which are finally joined are those minimizing the product of the corresponding cluster sizes. In the case illustrated, one would 
join the nodes $i_1$ and $j_1$, which yield a smaller product cluster size than $i_2$ and $j_2$
($2\cdot 5=10$ versus $3\cdot 4=12$).}
\label{fig:scheme}
\end{figure}

Let us first define an Achlioptas growth process.
The goal is to construct  a random  network of $N$ nodes and given 
degree sequence $\{k_1, k_2, \ldots , k_N \}$. If links are placed randomly, the procedure can 
be carried out with the configuration model~\cite{molloy95}.
Here instead, the criterion to add links is different.
At the beginning of the algorithm (i.e., stage $t=0$), we set $k_s(0)=k_s$
for each node $s$ (the only condition needed is that $\sum_s k_s$ should be an even number). 
The variables $k_s(t)$ act as a sort of counters: whenever a stub incident on node $i$ is connected to another stub
incident on node $j$, $k_i(t+1)=k_i(t)-1$ and $k_j(t+1)=k_j(t)-1$.
The construction proceeds until $T=\frac{1}{2}\sum_s k_s$ links have been drawn, which stands for $k_s(T)=0 \;, \, \forall \, s$
(i.e. there are no more stubs to be connected between node pairs). 
At each stage $t$ of the growth,
two pairs of vertices $\left(i_1,j_1\right)$ and $\left(i_2,j_2\right)$ are selected as candidate links:
these nodes are randomly selected among all vertices in the network with
probabilities $p_{i_1}(t) = k_{i_1} (t) / \sum_s k_s(t)$, $p_{j_1}(t) = k_{j_1}(t) / \sum_s k_s(t)$,
$p_{i_2}(t) = k_{i_2} (t) / \sum_s k_s(t)$ and $p_{j_2}(t) = k_{j_2} (t) / \sum_s k_s(t)$,
respectively, which basically means that the candidate links $\left(i_1,j_1\right)$ and 
$\left(i_2,j_2\right)$ are respectively selected with
probabilities $p_{\left(i_1,j_1\right)} (t) = p_{i_1}(t) \, p_{j_1}(t)$
and $p_{\left(i_2,j_2\right)} (t) = p_{i_2}(t) \, p_{j_2}(t)$.
In order to decide which of the two candidate links should be selected to
become a real link to be added to the network, one computes the quantities $L_{\left(i_1,j_1\right)}(t)=M_{i_1}(t)\,M_{j_1}(t)$ and
$L_{\left(i_2,j_2\right)}(t)=M_{i_2}(t)\,M_{j_2}(t)$, expressing the product of the sizes
of the clusters that the two selected links would merge (Fig.~\ref{fig:scheme}).
Finally, one draws the link for which the quantity $L$ is lower.
The former selection rule is called product rule (PR).
In principle other different reasonable criteria
may be used instead of the PR: taking the sum instead of the product, maximizing instead of
minimizing, etc.. 
During the construction of the network, one should avoid the presence of multiple links
(links connecting pairs of nodes already connected) and self-loops (links starting and ending at the same node).
Scale-free networks may have a significant number of multiple links and self-loops~\cite{boguna04}, but 
in the transition regime we are interested in here they are essentially tree-like (most links have to be
still placed), so multiple links and self-loops are very unlikely. In fact, we have verified that results do not change 
whether one allows or avoids them.

On Erd\"os-R\'enyi graphs the process we have described generates
a discontinuous percolation transition~\cite{achlioptas09}. More recently, 
Ziff has studied the same process for bond percolation on two-dimensional
square lattices~\cite{ziff09}, finding again a discontinuous transition.

A natural parameter which allows to follow the construction of the network is $p=t/T$, 
which expresses the fraction of links added to the network during its growth.
Following the construction of the network as a function of $p$ allows to study 
the formation of the giant component and the associated percolation 
transition of the network. This technique
allows to create the whole phase diagram of the transition through a single simulation~\cite{newman2000}.
\\
Let us define as order parameter the percolation strength $S^{(1)} = M^{(1)}/N$, where $M^{(1)}$
indicates the relative mass (i.e., number of nodes) belonging to the largest connected component in the network.
If the transition is continuous (i.e., second-order), the theory of finite size scaling tells us
that the percolation strength of a network composed of $N$ nodes obeys the relation
\begin{equation}
S^{(1)} = N^{-\beta/\nu} \, F\left[ \left( p-p_c \right) \, N^{1/\nu} \right],
\label{eq:perc_str}
\end{equation}
where $p_c$ is the percolation threshold (in the limit of
systems of infinite size), $\beta$ and $\nu$ are critical exponents of the transition
and $F(\cdot)$ is a universal function. Similar laws of finite size scaling may be written for
other observables. Here we consider the susceptibility
$\chi = N \sqrt{\langle {S^{(1)}}^2 \rangle - \langle S^{(1)} \rangle^2}$, which quantifies
the amplitude of the fluctuations of the percolation strength. The susceptibility $\chi$ obeys 
the relation 
\begin{equation}
\chi = N^{\gamma/\nu} \, G\left[ \left( p-p_c \right) \, N^{1/\nu} \right], 
\label{susc}
\end{equation}
where $\gamma$ is another critical exponent which characterizes the transition and $G(\cdot)$ is a universal function.
The susceptibility $\chi$ is directly related to the order parameter $S^{(1)}$. From 
the definition of $\chi$ and the scaling behavior of $S^{(1)}$ at $p_c$ (Eq.~\ref{eq:perc_str}), we deduce that $\gamma/\nu=1-\beta/\nu$.
\\
The susceptibility $\chi$ can be used for the determination of the critical point $p_c$.
The percolation threshold $p_c(N)$
of a system of finite size $N$ obeys the relation
\begin{equation}
p_c(N)=p_c+ b N^{-1/\nu} \;\;\;.
\label{eq:thres}
\end{equation}
$p_c(N)$ can be determined by finding the value of $p$ for which the
absolute maximum of $\chi$ occurs. Then a simple linear fit (based on the
maximization of the Pearson's correlation coefficient) of $p_c(N)$ vs. $N^{-1/\nu}$
allows to simultaneously compute both values of $p_c$ and $\nu$.
The same kind of analysis may be performed by determining $p_c(N)$
as the value of $p$ at which one observes the absolute maximum of $S^{(2)}$
(i.e., the relative size of the second largest component)~\cite{wu07}.
In our numerical simulations, we find a perfect agreement between the two different approaches.
\\
When the transition is discontinuous (i.e., first-order),
finite size scaling does not work. The scaling relations~(\ref{eq:perc_str}) and (\ref{susc}) 
trivially apply with $\beta/\nu=0$ and $\gamma/\nu=1$. 
The curves $S^{(1)}$ vs. $p$ corresponding to different system sizes do not scale
and $p_c(N)$ approaches $p_c$ faster than as a power law in the limit of large $N$.

We consider scale-free networks with degree exponent $\lambda$ [i.e., $P(k) \sim k^{-\lambda}$, where $P(k)$ is the
probability that a node has degree equal to $k$]. 
We examined two main scenarios, by setting the networks' cut-off (i.e., largest degree)
equal to $\sqrt{N}$ and to $N^{1/\left(\lambda-1\right)}$. The results however do not
qualitatively depend on this choice (the results shown refer to 
the cut-off $\sqrt{N}$). 
\\
\begin{figure}
\begin{center}
\includegraphics[width=0.47\textwidth]{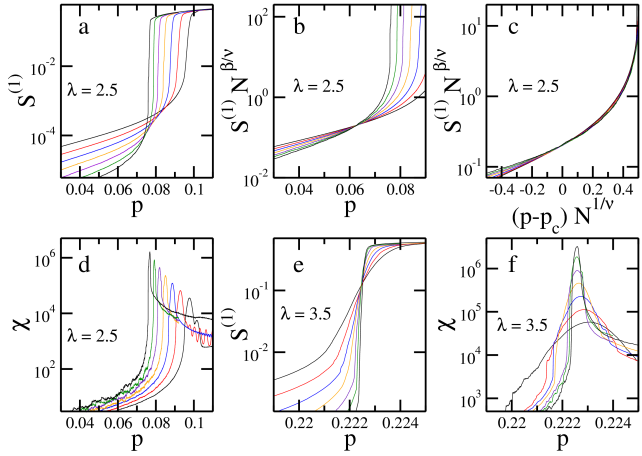}
\end{center}
\caption{Explosive percolation transition in scale-free networks. For $\lambda=2.5$ the transition
is continuous. In (a) and (b) we show the percolation strengths corresponding to 
different system sizes and their
rescaling $S^{(1)} N^{\beta/\nu}$, respectively.
The validity of Eq.(\ref{eq:perc_str}) can be proved by plotting
$S^{(1)} N^{\beta/\nu}$ vs. $\left(p-p_c\right) N^{1/\nu}$ (c). 
The peak of the susceptibility $\chi$ moves gradually
towards $p_c$ as the system size increases (d). Instead, for $\lambda=3.5$ the transition
is discontinuous: percolation strengths corresponding to different system sizes do not have a scaling form (e).
The location of the peaks of the susceptibility is essentially the same for any system size (f). The network sizes 
go from 256000 to 16384000, via successive doublings.}
\label{fig:scaling}
\end{figure}

When a scale-free network is constructed via an Achlioptas growth process,
the formation of the giant component is delayed. One needs to add a fraction of
links much larger than in a standard random process before seeing
the emergence of the giant component. Interestingly, for $\lambda<3$ it is
already possible to measure a non-vanishing value of the percolation threshold.
As an illustrative example, in Fig.~\ref{fig:scaling} we show the behavior of the order parameter $S^{(1)}$ 
and the susceptibility $\chi$ as a function of $p$.
We see that for $\lambda=2.5$ the scenario is the one expected for a continuous transition,
as confirmed by the scaling behavior of $S^{(1)}$ of Fig.~(\ref{fig:scaling}c), whereas for $\lambda=3.5$ the situation is different. 

\begin{figure}
\begin{center}
\includegraphics[width=0.47\textwidth]{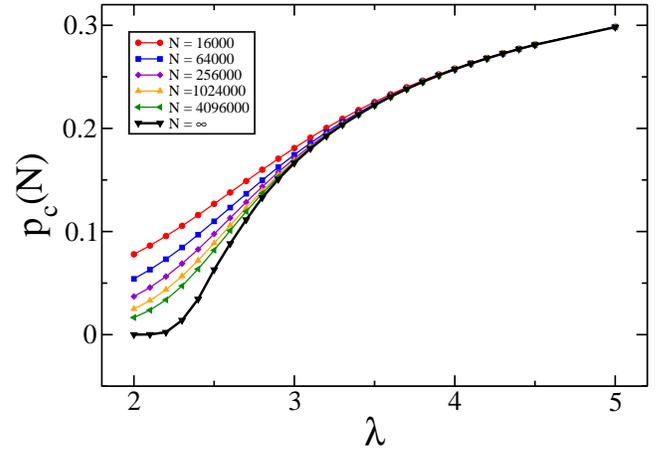}
\end{center}
\caption{Percolation threshold $p_c(N)$ as a function of the degree exponent $\lambda$ for various network sizes $N$. 
The black line represents the infinite size limit extrapolation  of the critical threshold. This extrapolation is made by 
using Eq.~(\ref{eq:thres}) for $\lambda \leq 3$, 
while for $\lambda \geq 3$ one uses the value of the plateau the pseudocritical points converge to.}
\label{fig:perc_threshold}
\end{figure}

We have carried out a detailed finite size scaling analysis of the percolation transition in the range of exponents 
$2\leq\lambda\leq 5$. For each value of $\lambda$ we have determined the pseudocritical 
point at a given system size $N$ and derived the infinite size limit of the threshold by using Eq.~(\ref{eq:thres}). 
In Fig.~\ref{fig:perc_threshold} we plot the lines of the pseudocritical points for various network sizes as a function of $\lambda$.
The black line indicates the extrapolation to the infinite size limit. The threshold is essentially zero up to 
$\lambda_c\sim 2.2$, and becomes non-zero for $\lambda>\lambda_c$. From our analysis 
we cannot exclude that for $2\leq\lambda\leq\lambda_c$ the threshold is non-zero but very small; 
in order to clarify the situation one should use systems of 
orders of magnitude larger than the ones we studied, which lie already at the boundary of what one could do without using supercomputers.

Interestingly, for $\lambda>3$ the pseudocritical point approaches
the actual threshold faster than as a power law, and the relation (\ref{eq:thres}) does not hold, 
which hints to a first order phase transition. We have confirmed the result by performing the test 
suggested by Achlioptas et al.~\cite{achlioptas09}.

\begin{figure}
\begin{center}
\includegraphics[width=\columnwidth]{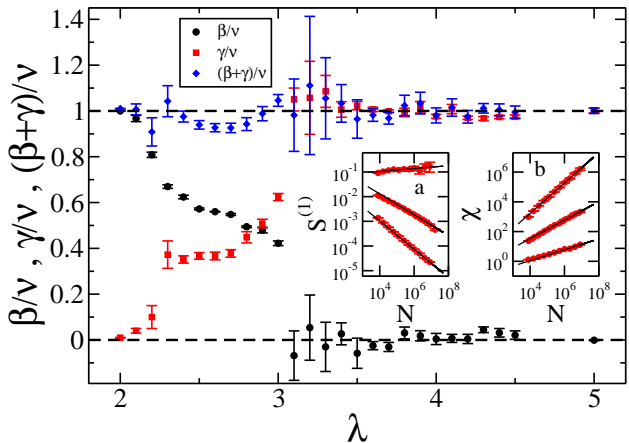}
\end{center}
\caption{Critical exponents' ratios $\beta/\nu$ and $\gamma/\nu$ as a function of the degree exponent $\lambda$. 
The insets show the scaling at $p_c$ of the quantities $S^{(1)} \sim N^{-\beta/\nu}$ (inset a) and $\chi \sim N^{\gamma/\nu}$ 
(inset b) for $\lambda=2.5, 3.0$ and $3.5$ (from bottom to top).}
\label{fig:perc_exponents}
\end{figure}

In the region of $\lambda$-values where we observe the second order phase transition we also computed the critical exponents,
by performing a finite size scaling analysis of the two main variables $S^{(1)}$ and $\chi$
at the critical point, according to Eqs.~(\ref{eq:perc_str}) and (\ref{susc}).
We have used such analysis also to double-check 
in an independent way the extrapolated values of the thresholds as a function of $\lambda$, which we had previously obtained 
from the scaling of Eq.~\ref{eq:thres}: the agreement is very good. 
The results are illustrated in Fig.~\ref{fig:perc_exponents}.
We plot the values of the exponents' ratios $\beta/\nu$, $\gamma/\nu$ and the sum $\beta/\nu+\gamma/\nu$.
We see that $\beta/\nu$, $\gamma/\nu$ are always in the range between $0$ and $1$, but their values 
depend on $\lambda$. The sum $\beta/\nu+\gamma/\nu$ is always $1$ with good approximation, as expected. We also remark that
around $\lambda_c$ the exponents display a jump. This is due to the fact that the threshold goes to very small values
for $\lambda<\lambda_c$ (consistent with zero), and finite size scaling cannot be accurate. For $\lambda>3$ the exponents
take trivial values: $\beta=0$, as the order parameter at criticality does not vanish in the infinite size limit; $\gamma=\nu$,
as the susceptibility is an extensive variable, as it should be if the transition were discontinuous. The insets show the
finite size scaling analysis on both exponents' ratios for three values of $\lambda$. For 
$S^{(1)}$ we see that, while for $\lambda<3$ there is
a clear power law scaling, as it should be for a continuous transition, for $\lambda>3$ there is a saturation. Similarly, for 
the susceptibility $\chi$, we see that the scaling is non-linear with $N$ for $\lambda<3$, whereas for  
$\lambda>3$ it becomes linear, as it happens for extensive quantities.

We have studied the percolation transition on static scale-free networks built with an Achlioptas process with 
product rule. We have found striking differences with standard percolation, from the existence of a finite 
threshold for $\lambda<3$ to the discontinuous character of the transition for $\lambda>3$. We stress that, since in an Achlioptas
process links are not placed completely at random, during the process the network generally has a different degree distribution,
and only at the end of the process, when all links are placed, one restores the original imposed distribution~\cite{cho09}. 
We have verified that
the networks at the percolation transition still have a power law degree distribution, but with a different exponent than the imposed one.
In particular, we have verified that $\lambda=2.2$ corresponds to the effective exponent $\lambda^\prime=3$.
This may explain the existence of a finite threshold for $\lambda>2.2$ (it would correspond to $\lambda^\prime>3$ for the actual
networks at the threshold), but not the origin of the discontinuous transition, which remains yet to be uncovered.

Our findings show that 
the building mechanism of scale-free networks may strongly affect dynamic processes taking place on the network, along with
structural features (e. g., resilience to failures/attacks), even if the degree distribution is predefined. So, very different
phenomena can occur on networks with exactly the same degree distribution. The process we have studied here
deserves further investigations, both from the numerical and the analytical point of view and it may reveal new exciting perspectives
in the field of complex networks and in the theory of critical phenomena. 
Moreover, this finding may open new perspectives in other fields where networks
are important, such as computer science and engineering. 
In particular, the issues of robustness and information transmission are inextricably linked
to percolation. 

We are indebted to J. J. Ramasco for bringing this problem to our attention. S. F. gratefully acknowledges ICTeCollective,
grant number 238597 of the European Commission. At the moment of submission of this manuscript
we have noticed a paper by Cho et al., just posted on the electronic archive~\cite{cho09}. The paper deals with the same problem, 
but the model used to build the network is not the same, which leads 
to significant discrepancies in the results. We apologize with Cho et al. for this unlucky and unwanted
coincidence.


\end{document}